\documentclass[smallextended]{svjour3}       
\smartqed  
\usepackage{graphicx}

\begin{document}

\title{Adsorption of Helium atoms on two-dimensional substrates \thanks{The work is performed according to the Russian Government
	Program of Competitive Growth of Kazan Federal
	University}
}


\author{Regina Burganova         \and
        Yury Lysogorskiy  \and    
            Oleg Nedopekin  \and 
               Dmitrii Tayurskii 
}


\institute{R. Burganova, Y. Lysogorskiy, O. Nedopekin, D. Tayurskii \at
             Institute of Physics, Kazan Federal University, Kremlevskaya St. 18, 420008 Kazan, Russia \\
             \email{bur.regina@gmail.com}                   
}

\date{Received: date / Accepted: date}

\maketitle

\begin{abstract} 
	The study of the adsorption phenomenon of helium began many decades ago with the discovery of graphite as a homogeneous substrate for investigation of physically adsorbed monolayer films. In particular, helium monoatomic layers on graphite were found to exhibit a very rich phase diagram.
	
	 In the present work we have investigated the adsorption phenomenon of helium atoms on graphene and silicene substrates by means of density functional theory with Born-Oppenheimer approximation. Helium-substrate and helium-helium interactions were considered from first principles. Vibrational properties of adsorbed monolayers have been used to explore the stability of the system. This approach reproduces results describing the stability of a helium monolayer on graphene calculated by quantum Monte Carlo (QMC) simulations for low and high coverage cases. However, for the moderate coverage value there is discrepancy with QMC results due to the lack of helium zero point motion.      
\end{abstract}

\section{Introduction}
\label{intro}

Since the discovery of graphite as a homogeneous substrate for investigation of physisorbed monolayer films and various helium phases on it~\cite{schick1980phase}, the most interesting question is whether He-He or He-C atomic interactions define helium behavior. 
The study of adsorption of helium atoms on solid substrates therefore is important for both adsorption properties of surfaces and behavior of He atoms. Moreover, it was found that restricted geometry, such as nanoporous media or a two-dimensional substrate, provides new and unique features of helium. For example, it was shown that in a porous medium a new helium phase appears~\cite{yamamoto2008thermodynamic}. Novel materials such as substrates can also lead to a new behavior of helium due to a different potential created by the substrate~\cite{nava2012adsorption}. 

It was shown that at certain densities and temperatures helium monolayers on graphene and graphite form commensurate and incommensurate solid phases and create a two-dimensional lattice~\cite{manchester1967solid,bretz1972solid}. Such structures were investigated previously~\cite{pierce1999path,gordillo1998path,gordillo2009he,kwon20124} by Quantum Monte Carlo (QMC) simulations. In the present work we have investigated the adsorption and stability of the first helium layers adsorbed on graphene by means of quasi harmonic approximation within density functional theory (DFT)~\cite{hohenberg1964,kohn1965self} in order to examine the method. Investigations were then carried out on the new two-dimensional material's surface silicene in order to discover new possible adsorption properties of helium.

Silicene is a graphene-like Si-based material, which was first predicted theoretically~\cite{guzman2007electronic} and then obtained experimentally~\cite{lalmi2010epitaxial}. The electronic structure of this material is almost identical to graphene~\cite{guzman2007electronic} but it exhibits a buckled surface and larger interatomic distance. One can expect, therefore, that similar solid phases of helium on silicene will be expanded. 

Simulations of helium adsorption in different media such as a graphene nanotube~\cite{timerkaeva2011ab} and a porous material called aerogel~\cite{debras2011ab,lysogorskiy2013density} within different variations of DFT methods have recently been obtained. All the results show that DFT methods give an accurate description of the adsorption phenomenon. Moreover, it is known that among \textit{ab initio} methods DFT allows a balance to be achieved between computational cost and accuracy.

\section{Simulation details}
\label{sec:1}

We started with the He-graphene system as a reference for verification of our simulation technique as there are a lot of experimental and theoretical data about adsorption and interaction of He atoms with graphene and graphite substrates~\cite{cole1981probing}. In addition, graphene and silicene both have similar structure and C and Si atoms are isovalent. 

The calculations were performed using DFT method with gradient-corrected exchange, correlation energy functionals and projector augmented-wave method~\cite{blochl1994projector,kresse1999ultrasoft} implemented in VASP code~\cite{kresse1996efficient} in MedeA \cite{medea} program. 

The maximum kinetic energy of plane waves in a basis set was equal to 480\,eV. To integrate in the Brillouin zone we used equidistant Mokhorst-Pack k-mesh $3 \times3 \times 1$ centered at Gamma point and Methfessel-Paxton \cite{methfessel1989high} smearing with the parameter 0.2\,eV. Optimization of the atomic positions was done by the conjugate gradient method with the maximal force equal to 0.005\,eV/\AA. In order to simulate the two-dimensional structure we added a 15\,\AA~vacuum slab in the perpendicular to the surface direction. We tested different computational approaches consisting of employing different types of exchange-correlation parts in the energy functional in order to better reproduce the He-graphene interaction potential. We also checked the van der Waals corrections due to the fact that at large distances, the He-substrate interaction is governed by van der Waals interaction. 
 
The obtained data was compared with the potential calculated by semiempirical method, based on the scattering of helium on graphite suggested in~Ref.\cite{Carlos1980a}.
\begin{figure}
	\centering
	\includegraphics[width=0.75\textwidth]{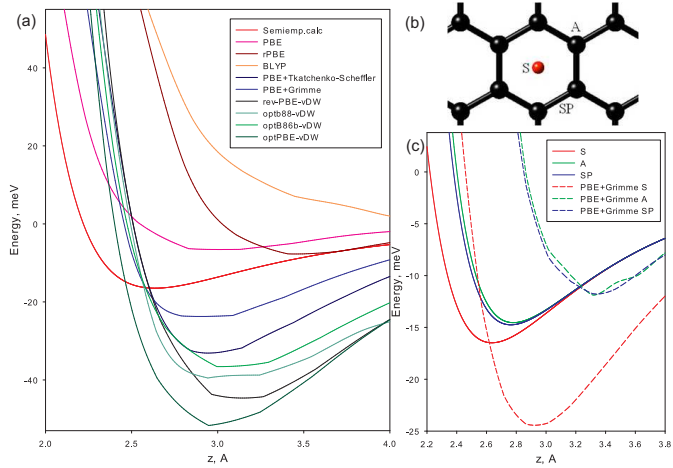}
	\caption{ Adsorption potentials of He atom on graphene: semiempirical~\cite{Carlos1980} and ab initio calculations using various functionals: GGA-PBE~\cite{perdew1996generalized}, GGA-rPBE~\cite{hammer1999improved}, BLYP~\cite{becke1988density}, GGA-PBE with Grimme-D2~\cite{grimme2006semiempirical} and Tkatchenko-Scheffler~\cite{tkatchenko2009accurate} van Der Waals corrections and functionals~\cite{klimevs2010chemical,dion2004van} (a). Various symmetrical points of He atom above graphene surface (b) and corresponding adsorption potentials (c). }
	\label{fig:Gr-He_interaction}     
\end{figure}
Taking into account the  van der Waals corrections~\cite{grimme2006semiempirical,tkatchenko2009accurate} is essential for correct reproduction of He-graphene interaction (see Fig.\ref{fig:Gr-He_interaction}). However, using such van der Waals functionals as rev-PBE~\cite{dion2004van}, optb88, optB86b~\cite{klimevs2010chemical} and optPBE  leads to overestimation of potential well. GGA-PBE functional with semiempirical DFT-D2~\cite{grimme2006semiempirical} correction for long-range interaction reproduced the semiempirical data well and was chosen for further simulations.

To investigate the stability of the adsorbed helium monolayers we have calculated vibrational properties using MedeA-PHONON module, which implements a direct approach of harmonic approximation~\cite{parlinski1997first}. The so-called direct approach to lattice dynamics is based on the ab initio evaluation of forces on all atoms produced by a set of finite displacements of a few atoms within an otherwise perfect crystal.

The simulation model of silicene consists of one atomic layer of silicone atoms with crystallographic surface (001) of bulk silicone and space group symmetry $P63mc$. The optimized hexagonal cell has a lattice size 3.85\,\AA and interatomic distance $d_\mathrm{Si-Si}= 2.26$\,\AA~ (close to $1.9\pm0.1$~\cite{lalmi2010epitaxial} and  $2.2\pm0.1$\,\AA~\cite{vogt2012silicene})and a buckling height  $\Delta=0.44$\,\AA. 

\section{Adsorption potential}
\label{sec:3}

We have calculated He-silicene interaction potentials in the perpendicular direction at various symmetric positions of He atom above the silicene surface. 

\begin{figure*}
	\centering
	\includegraphics[width=0.73\textwidth]{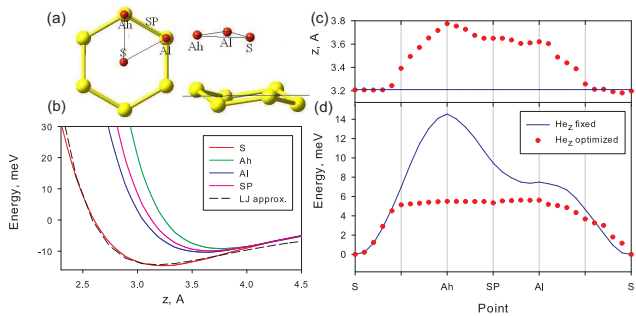}
	\caption{ Various symmetrical positions of He atom above silicene surface (a) and corresponding adsorption potentials as well as approximation of adsorption curve as a sum of LJ pairwise interactions above S point (b). The equilibrium height (c) and potential energy profile (d) along the path through symmetrical positions of He atom.}
	\label{fig:Si-He_interaction}     
\end{figure*}

Fig.~\ref{fig:Si-He_interaction}(b) demonstrates that the deepest potential well for helium is on the top of point S, corresponding to the center of the Si hexagon. The depth of the potential well for a helium atom is $U_{0}$ = 14\,meV and equilibrium height is $z_{0}$ = 3.20\,\AA. The positions Ah, Al and SP of a helium atom have energy of about 4\,meV higher, which corresponds to 48\,K. Thus the position S is the most preferable adsorption site at low temperatures.  The interaction potential above the S point was approximated as a sum of pairwise Lennard-Jones (LJ) interactions between He and each Si atom in a range 10\,\AA. LJ parameters are $\epsilon = 1.62$\,meV and $\sigma =$ 3.6\,\AA.  We have computed a potential energy profile of a helium atom on the silicene surface along the lines shown in Fig.~\ref{fig:Si-He_interaction}(a). It can be seen in Fig.~\ref{fig:Si-He_interaction}(d) that the center of the hexagon (S point) represents a potential well for the helium atom as in the perpendicular direction.

\section{Results and discussion}
\label{sec:4}
\begin{figure}
	\centering
	\includegraphics[width=0.70\textwidth]{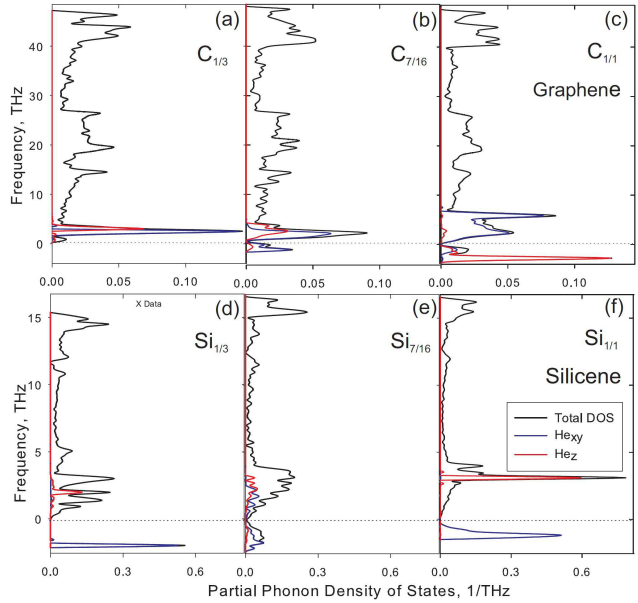}
	\caption{Vibrational density of states for the C$_{1/3}$ (a),  C$_{7/16}$ (b) and C$_{1/1}$ (c) helium phases on graphene and  Si$_{1/3}$ (d),   Si$_{7/16}$ (e) and Si$_{1/1}$ (f) helium phases on silicene. Presence of imaginary modes indicates structural instability.}   
	\label{fig:Gr-He_DOS}  
\end{figure}
Recent QMC calculations have shown that a helium monolayer adsorded on graphene at low temperatures could demonstrate different phases depending on helium coverage ~\cite{happacher2013phase}. At low coverages there is stable C$_{1/3}$ commensurate solid phase, which is characterized by one He atom corresponding to three adsorption sites (S points)~\cite{happacher2013phase}. At higher helium coverage range the domain walls phase exists. At a certain He density value the C$_{7/16}$ commensurate solid phase occurs. With subsequent density increase, the incommensurate phases are formed. At even higher helium coverage a second He layer should appear~\cite{happacher2013phase}.

We have investigated stability of an adsorbed helium monolayer on graphene in three different phases C$_{1/3}$, C$_{7/16}$ and C$_{1/1}$ (with 0.063, 0.083, 0.19\,\AA$^{-2}$ coverages respectively). The presence of vibrational modes with imaginary frequencies usually point out the structural instability. In Fig~\ref{fig:Gr-He_DOS}  the density of states of vibrational spectra are given. As one can see, phase C$_{1/3}$ does not have any imaginary modes (see Fig.~\ref{fig:Gr-He_DOS} (a)), implying stability of adsorbed layer, whereas C$_{7/16}$ and C$_{1/1}$ phases are unstable (Fig.~\ref{fig:Gr-He_DOS}(b) and (c)). However, QMC simulation demonstrates the stability of the C$_{7/16}$ phase. This discrepancy could be explained by the lack of He atom zero point motion in our approach, where this effect could stabilize the inplane motion. 
Imaginary modes for vibrations of helium in the C$_{1/1}$ phase along $z$ direction indicates layer promotion because of high He-He atoms repulsion, which is agrees well agreed with QMC calculations.                  

In the case of helium on silicene the same phases correspond to 0.026, 0.034, 0.078\,\AA$^{-2}$ coverages respectively. The simulations show  its instability by the presence of imaginary modes for He atoms in $xy$ plane (Fig.~\ref{fig:Gr-He_DOS}(d-f)).  It could be interpreted as the tendency of He atoms to group into small clusters with higher density because of the large Si-Si distance and less attractive He-silicene potential in comparison to graphene.

\section{Conclusions}
\label{sec:5}

We have explored stability of adsorbed monoatomic layers of helium within Born-Oppenheimer (BO) approximation for graphene substrate and found that for low coverages there is conformity with QMC calculations. As coverage is increased zero point effects become dominant and results of BO and QMC simulations differ. However, BO calculation points to the formation of a second layer for high He coverage that correlates with QMC calculations. 

The similar phases of helium on silicene substrate were found to be unstable.  The depth of the potential well for He atom on silicene is about 14\,meV  compared to 24\,meV on graphene according to our calculations. The equilibrium position of a He atom above these surfaces also differs -- 3.20\,\AA~and 2.93\,\AA~for silicene and graphene respectively. One can conclude, therefore,  that silicene is less attractive substrate for He atom adsorption than graphene.


\end{document}